\documentclass[prd,12pt,a4paper,showpacs,nofootinbib,superscriptaddress]{revtex4-2}
\usepackage{bm}
\usepackage{indentfirst}
\usepackage{amsmath}
\usepackage{amssymb}
\usepackage{graphicx}
\usepackage{float}
\usepackage{subfigure}
\usepackage{hyperref}
\hypersetup{
	colorlinks=true,
	linkcolor=red,
	citecolor=blue,
}
\begin{document}
\title{Detection of gravitational wave mixed polarizations with single space-based detectors} 	

\author{Chao Zhang}
\email{chao\_zhang@hust.edu.cn}
\affiliation{School of Physics, Huazhong University of Science and Technology,
Wuhan, Hubei 430074, China}

\author{Yungui Gong}
\email{Corresponding author. yggong@hust.edu.cn}
\affiliation{School of Physics, Huazhong University of Science and Technology, Wuhan, Hubei 430074, China}

\author{Dicong Liang}
\email{dcliang@pku.edu.cn}
\affiliation{Kavli Institute for Astronomy and Astrophysics, Peking University, Beijing
100871, China}
\affiliation{School of Physics, Huazhong University of Science and Technology,
Wuhan, Hubei 430074, China}

\author{Chunyu Zhang}
\email{chunyuzhang@hust.edu.cn}
\affiliation{School of Physics, Huazhong University of Science and Technology,
Wuhan, Hubei 430074, China}

\begin{abstract}
General Relativity predicts only two tensor polarization modes for gravitational waves while at most six possible polarization modes are allowed in general metric theory of gravity.
The number of polarization modes is determined by the specific modified theory of gravity. Therefore, the determination of polarization modes can be used to test gravitational theory.
We introduce a concrete data analysis pipeline for a space-based detector such as LISA to detect the polarization modes of gravitational waves.
This method can be used for monochromatic gravitational waves emitted from any compact binary system with known sky position and frequency to detect mixtures of tensor and extra polarization modes.
We use the source J0806.3+1527 with one-year simulation data as an example to show that this approach is capable of probing pure and mixed polarizations without knowing the exact polarization modes.
We also find that the ability of detection of extra polarization depends on the gravitational wave source location and the amplitude of non-tensorial components.
\end{abstract}

\maketitle
\section{Introduction}
So far there have been tens of confirmed gravitational wave (GW) detections  \cite{Abbott:2016blz,TheLIGOScientific:2016agk,Abbott:2016nmj,Abbott:2017vtc,Abbott:2017oio,TheLIGOScientific:2017qsa,Abbott:2017gyy,LIGOScientific:2018mvr,Abbott:2020uma,LIGOScientific:2020stg,Abbott:2020khf,Abbott:2020tfl,Abbott:2020niy,LIGOScientific:2021usb} since the first GW event GW150914 observed
by the Laser Interferometer Gravitational-Wave Observatory (LIGO) Scientific Collaboration and the Virgo Collaboration \cite{Abbott:2016blz,TheLIGOScientific:2016agk}.
Distinguishing GW polarizations is extremely useful to perform test about the validity of General Relativity (GR).
The transient GWs detected by ground-based GW detectors are the merging signals with the duration of seconds to minutes in the frequency band around several-hundred hertz,
so it is impossible to measure the signals' polarization contents with advanced LIGO alone because the two detectors are nearly co-oriented \cite{Harry:2010zz,TheLIGOScientific:2014jea}
and the observed signals are so short that we can ignore the motion of the detector around the Sun.
However, some preliminary results on the signals' polarization contents were obtained with the LIGO-Virgo network \cite{Abbott:2017oio,LIGOScientific:2018mvr,Abbott:2020niy,Abbott:2018lct}.
In general metric theory of gravity, GWs can have up to six polarization modes \cite{Eardley:1973br,Eardley:1974nw}: two transverse-traceless tensor modes  ($+$ and $\times$), two vector modes ($x$ and $y$), a scalar breathing mode ($b$) and a scalar longitudinal mode ($l$).
The specific modified theory of gravity uniquely determines the polarization modes.
For example, in Brans-Dicke theory \cite{Brans:1961sx} there exists one extra breathing mode beyond the two transverse-traceless tensor modes of GR. The scalar polarization mode is a mixture of breathing mode and longitudinal mode if the scalar field is massive in the generic scalar-tensor theory of gravity \cite{Liang:2017ahj,Hou:2017bqj,Gong:2018vbo,Gong:2018cgj}.
Einstein-{\AE}ther theory \cite{Jacobson:2004ts} predicts the existence of scalar and vector polarization modes \cite{Lin:2018ken,Zhang:2019iim,Gong:2018cgj} while generalized tensor-vector-scalar theories, such as TeVeS theory \cite{Bekenstein:2004ne}, predict the existence of all 6 polarization modes \cite{Gong:2018cgj}.
Therefore, the detection of extra polarization modes allows us to falsify GR.
To separate the polarization modes of GWs, in principle the number of ground-based GW detectors oriented differently should be equal to or larger than the number of the polarization modes.
The network of ground-based GW detectors including advanced LIGO \cite{Harry:2010zz,TheLIGOScientific:2014jea}, advanced Virgo \cite{TheVirgo:2014hva}, KAGRA \cite{Somiya:2011np,Aso:2013eba} and LIGO India has the ability of probing extra polarization modes \cite{Hagihara:2018azu,Hagihara:2019ihn,Pang:2020pfz}.
In the past years, different methods were developed to probe nontensorial polarizations
in stochastic GW backgrounds \cite{Nishizawa:2009bf,Callister:2017ocg,Abbott:2018utx,Nishizawa:2009jh}, continuous GWs \cite{Isi:2015cva,Isi:2017equ,Abbott:2017tlp,OBeirne:2019lwp},
GW bursts \cite{Hayama:2012au,DiPalma:2017qlq} and GWs from compact binary coalescences \cite{Takeda:2018uai,Takeda:2019gwk}.
In particular, the Fisher information matrix approximation was usually used to estimate the parameters of the source and to discuss the measurement of polarization modes \cite{Takeda:2019gwk,Vallisneri:2007ev,Wen:2010cr,Aasi:2013wya,Grover:2013sha,Berry:2014jja,Singer:2015ema,Becsy:2016ofp,Zhao:2017cbb,Mills:2017urp,Fairhurst:2017mvj,Fujii:2019hdi,Liu:2020mab,Zhang:2020hyx,Zhang:2020drf,Zhang:2021kkh,Gong:2021gvw}.

For stellar or intermediate black hole binaries with the mass range $100-10^4~M_{\odot}$, in the early inspiral phase the GW frequency is in the mHz range and its evolution can be neglected during the mission of the space-based GW detector.
The proposed space-based GW observatories including LISA \cite{Danzmann:1997hm,Audley:2017drz},
TianQin \cite{Luo:2015ght} and Taiji \cite{Hu:2017mde}
can detect the monochromatic GW signals emitted by these wealthy sources.
Due to the orbital motion of the detector in space,
along its trajectory the single detector can be effectively regarded as a set of virtual detectors and therefore form a virtual network to measure the polarization contents of the monochromatic GW signals.
By taking a specific linear combination of the outputs in the network of detectors it is possible to remove any tensorial signal present in the data  \cite{Guersel:1989th,Chatterji:2006nh}.
A particular $\chi^2$ distribution is followed by the null energy constructed with this method \cite{Chatterji:2006nh} when the null energy is calculated at
the true sky position.
If nontensorial polarization exists in the data,
then the null energy evaluated at the true sky position no longer follows the particular $\chi^2$ distribution \cite{Pang:2020pfz}.
Based on these results, we introduce one concrete data analysis pipeline to check the existence of extra polarization for a single space-based GW detector.
Apart from being able to detect mixtures of tensor polarization modes and alternative polarization modes,
this method has the added advantage that no waveform model is needed,
and monochromatic GWs from any kind of compact binary systems with known sky positions and frequencies can be used.

The paper is organized as follows.
In Section \ref{sec2}, we describe the basics of the GW signal registered in the space-based GW detector.
In Section \ref{sec:formulation}, we present general monochromatic waveforms including extra polarization modes
and construct the method to discover alternative polarization modes.
We then apply the method on the source J0806.3+1527 with one-year simulation data for LISA, Taiji, and TianQin.
Our conclusion and discussion are presented in Section \ref{sec:final}.

\section{Gravitational Wave Signal}
\label{sec2}
It is convenient to describe GWs and the motion of space-based GW detectors like LISA, TianQin, and Taiji in the heliocentric coordinate system with the constant basis vectors $\left\{\hat{e}_x,\hat{e}_y,\hat{e}_z\right\}$ \cite{Rubbo:2003ap}.
For GWs propagating in the direction $\hat{\omega}$,
we introduce a set of unit vectors $\{\hat{\theta}, \hat{\phi}, \hat{\omega}\}$ which are perpendicular to each other,
\begin{equation}\label{polarization}
\begin{aligned}
\hat{\theta}&=\cos(\theta)\cos(\phi)\hat{e}_x+ \cos(\theta)\sin(\phi) \hat{e}_y-\sin(\theta)\hat{e}_z,\\
\hat{\phi}&=-\sin(\phi)\hat{e}_x+\cos(\phi)\hat{e}_y,\\
\hat{\omega} &=-\sin(\theta)\cos(\phi)\hat{e}_x -\sin(\theta)\sin(\phi)\hat{e}_y -\cos(\theta)\hat{e}_z,
 \end{aligned}
\end{equation}
  where the angles $(\theta,\phi)$ are the angular coordinates of the source.
  To describe the six possible polarization modes of GWs in general metric theory of gravity, the polarization angle $\psi$ is introduced to form polarization axes of the gravitational radiation,
\begin{equation}
\label{polcoor1}
\hat{p}=\cos\psi \hat{\theta}+\sin \psi \hat{\phi},\quad \hat{q}=-\sin\psi \hat{\theta}+\cos\psi \hat{\phi}.
\end{equation}
The polarization tensors are
\begin{equation}
\begin{split}
e^{+}_{ij}=\hat{p}_i\hat{p}_j-\hat{q}_i\hat{q}_j, \quad & e^{\times}_{ij}=\hat{p}_i\hat{q}_j+\hat{q}_i\hat{p}_j,\\
e^{x}_{ij}=-\hat{p}_i\hat{\omega}_j-\hat{\omega}_i\hat{p}_j, \quad & e^{y}_{ij}=-\hat{q}_i\hat{\omega}_j-\hat{\omega}_i\hat{q}_j,\\
e^{l}_{ij}=\sqrt{2}\hat{\omega}_i\hat{\omega}_j, \quad & e^{b}_{ij}=\hat{p}_i\hat{p}_j+\hat{q}_i\hat{q}_j,
\end{split}
\end{equation}
where $+$ and $\times$ denote two transverse-traceless tensor modes,
$x$ and $y$ denote two vector modes, $b$ denotes the scalar breathing mode and $l$ denotes the scalar longitudinal mode.
In terms of the six polarization tensors $e^A_{ij}$, GWs in general metric theory of gravity have the form
\begin{equation}
h_{ij}(t)=\sum_{A}e^A_{ij}h_A(t),
\end{equation}
where $A=+,\times,x,y,l,b$.

For a monochromatic GW with the frequency $f$ propagating along the direction $\hat{\omega}$, arriving at the Sun at a time $t$,
the output in an equal-arm space-based interferometric detector such as LISA, TianQin and Taiji  with a single round-trip of light travel  is
\begin{equation}
\label{gwst}
s(t)=\sum_A F^A h_A(t)e^{i\phi_D(t)},
\end{equation}
\begin{equation}
\label{doppler}
   \phi_D(t)=\frac{2\pi  f R}{c}\sin\theta\cos\left(\frac{2\pi t}{P}+\phi_{\alpha} -\phi\right),
\end{equation}
\begin{equation}
\label{poltensor1}
F^A=\sum_{i,j} D^{ij} e^A_{ij},
\end{equation}
where $F^A$ is the pattern function for the polarization mode A, $\phi_D(t)$ is the Doppler phase, $\phi_{\alpha}$ is the ecliptic longitude of the detector $\alpha$ at $t=0$,
the rotational period $P$ is 1 yr and the radius $R$ of the orbit is 1 A.U. 
The detector tensor $D^{ij}$  is
\begin{equation}
D^{ij}=\frac{1}{2}\left[\hat{u}^i \hat{u}^j T\left(f,\hat{u}\cdot\hat{\omega}\right)-\hat{v}^i \hat{v}^j T\left(f,\hat{v}\cdot\hat{\omega}\right)\right],
\end{equation}
where $\hat{u}$ and $\hat{v}$ are the unit vectors along the arms of the detector.
The detailed orbit equations are presented in Appendix~\ref{orbits}.
Additionally, $T(f,\hat{u}\cdot\hat{\omega})$ is \cite{Estabrook:1975,Cornish:2001qi}
\begin{equation}
\label{transferfunction}
\begin{split}
T(f,x)=\frac{1}{2}&
\left\{\text{sinc}\left[\frac{f(1-x)}{2f^*}\right]\exp\left[\frac{f(3+x)}{2if^*}\right] \right.\\
&\left. +\text{sinc}\left[\frac{f(1+x)}{2f^*}\right]\exp\left[\frac{f(1+x)}{2if^*}\right]\right\},
\end{split}
\end{equation}
where $\text{sinc}(x)=\sin x/x$, $f^*=c/(2\pi L)$ is the transfer frequency of the detector, $c$ is the speed of light and $L$ is the arm length of the detector.
Note that in the long wavelength approximation $f\ll f^*$,
we have $T(f, \hat{u}\cdot\hat{\omega})\rightarrow$ 1 in Eq. (\ref{transferfunction}).
The triangle configuration of the proposed space-based GW detector such as LISA, TianQin and Taiji can be  regarded as two L-shaped detectors effectively.
In this paper we only consider the Michelson interferometer consisting of two equal arms with the unit vectors $\hat{u}$ and $\hat{v}$ for simplicity.
To significantly reduce the laser frequency noise due to unequal arm lengths, 
time-delay interferometry (TDI) \cite{Tinto:1999yr,Armstrong_1999} is needed. 
We discuss the GW response for the TDI Michelson variable $X$ in Appendix \ref{tdigw}.
The averaged response function $|F^A|^2$ including different TDI combinations
for different polarization mode $A$ was discussed in  \cite{Larson:1999we,Larson:2002xr,Tinto:2010hz,Blaut:2012zz,Liang:2019pry,Zhang:2019oet,Zhang:2020khm}.

\section{Methodology}
\label{sec:formulation}
Now we consider the strain output $d(t)$ produced by a monochromatic GW for a space-based GW detector in the heliocentric coordinate system.
A monochromatic GW  assumed to be emitted from a source with the sky location $-\hat{\omega} (\theta, \phi)$, arrives
at the Sun at the time $t$.
If only the tensor polarization modes are present, we have
\begin{equation}
\label{signal_tensor}
\begin{split}
d_w(t)= F_w^{+}&(\hat{\omega},f,t)h_+(t)e^{i\phi_D(t)}\\ &+F_w^{\times}(\hat{\omega},f,t)h_\times(t)e^{i\phi_D(t)}
+ n_w(t),
\end{split}
\end{equation}
where $F_w^{+}$ and $F_w^{\times}$ are the noise-weighted beam pattern functions and $n_w(t)$ is the whitened noise.
The noise-weighted beam pattern functions and noise-weighted data are \cite{Allen:2005fk}
\begin{equation}
F_{w}^A=\frac{F^A}{\sqrt{S_n(f)}},\quad d_{w}=\frac{d}{\sqrt{S_n(f)}}.
\end{equation}
In the following we always use the noise-weighted beam pattern functions and noise-weighted data,
so we ignore the label $w$ for simplicity.
For space-based interferometers, the noise power spectral density $S_n(f)$ is \cite{Cornish:2018dyw,Luo:2015ght,Hu:2018yqb,Audley:2017drz}
\begin{equation}\label{Sn}
S_n(f)=\frac{S_x}{L^2}+\frac{2S_a\left(1+\cos^2(f/f^*) \right)}{(2\pi f)^4L^2}\left(1+\left(\frac{0.4~\text{mHz}}{f}\right)^2\right).
\end{equation}
For LISA, the acceleration noise is $\sqrt{S_a}=3\times 10^{-15}\ \text{m s}^{-2}/\text{Hz}^{1/2}$,
the displacement noise is $\sqrt{S_x}=15\ \text{pm/Hz}^{1/2}$,
the arm length is $L=2.5 \times 10^6$ km,
and its transfer frequency is $f^*=0.02$ Hz \cite{Audley:2017drz}.
Similarly, for TianQin $\sqrt{S_a}=10^{-15}\
\text{m s}^{-2}/\text{Hz}^{1/2}$,
$\sqrt{S_x}=1\ \text{pm/Hz}^{1/2}$, $L=\sqrt{3} \times 10^5$ km,
and $f^*=0.28$ Hz \cite{Luo:2015ght}.
For Taiji $\sqrt{S_a}=3\times10^{-15}\
\text{m s}^{-2}/\text{Hz}^{1/2}$,
$\sqrt{S_x}=8\ \text{pm/Hz}^{1/2}$, $L=3\times 10^6$ km,
and $f^*=0.016$ Hz \cite{Guo:2018npi}.

Taking the source J0806.3+1527 located at $(\theta=94.7^\circ, \phi=120.5^\circ)$ as an example, we
simulate the strain output in a space-based GW detector.
In GR the quadrupole formula provides
the lowest-order post-Newtonian GW waveform for a binary system as \footnote{In Eq. \eqref{hpceq1}, $f_0$ should be the observed frequency which is related with the emitted frequency $f_e$ as $f_0=f_e/(1+z)$ and the chirp mass $\mathcal{M}$ in the source frame should be $(1+z)\mathcal{M}$ in the detector frame.}
\begin{equation}
\label{hpceq1}
\begin{split}
h_+=&\mathcal{A}\left[1+\cos^2(\iota)\right]\exp (2\pi i f_0 t + i\phi_0),\\
h_\times=&2i\mathcal{A}\cos(\iota)\exp (2\pi i f_0 t + i\phi_0),
\end{split}
\end{equation}
where $\mathcal{A}=2(G\mathcal{M}/c^2)^{5/3}(\pi f/c)^{2/3}/D_L$  is the GW overall amplitude, $\mathcal{M} = 0.3 ~M_{\odot}$ (we take the component masses
$0.5 ~M_{\odot}$ and $0.25 ~M_{\odot}$) is the chirp mass,
$D_L = 0.5$ kpc is the luminosity distance,
$\iota = \pi/6$ is the inclination angle between the line of sight
and the binary orbital axis, $\phi _0$ is the initial GW phase at the start of observation,
$f_0=6.22 ~\rm{mHz}$ is the emitted GW frequency of the source J0806.3+1527  \cite{Israel:2002gq,Barros:2004er,Roelofs:2010uv,Esposito:2013vja,Kupfer:2018jee}.
The signal in Eq.~\eqref{signal_tensor} can be rewritten in another form
\begin{equation}
\begin{split}
d(t)=\bar{h}_+ &F^{+}(\hat{\omega},f_0,t) e^{2\pi if_0 t+i\phi_D(t)} \\
&+ \bar{h}_\times F^{\times}(\hat{\omega},f_0,t) e^{2\pi if_0 t+i\phi_D(t)}
+ n(t),
\end{split}
\end{equation}
where
\begin{align*}
\bar{h}_+=&\mathcal{A}\left[1+\cos^2(\iota)\right]\exp ( i\phi_0), \\
\bar{h}_\times=&2i\mathcal{A}\cos(\iota)\exp (i\phi_0).
\end{align*}
We denote $N$ number of the  observational data $d[k]=d(t_k)$ at discrete times in a more compact matrix form
\begin{equation}
\mathbf{d}[k] = \mathbf{Fh}[k]+\mathbf{n}[k],
\label{eq:d_Fh}
\end{equation}
where
\begin{equation}
\mathbf{d}=
\begin{pmatrix}
d[0] \\
\vdots \\
d[k]
\end{pmatrix}
\text{,\,\,\,\,\,\,\,}
\mathbf{h}=
\begin{pmatrix}
\bar{h}_{+} \\
\bar{h}_{\times}
\end{pmatrix}
\text{,\,\,\,\,\,\,\,}
\mathbf{n}=
\begin{pmatrix}
n[0] \\
\vdots \\
n[k]
\end{pmatrix},
\end{equation}
\begin{equation}
\begin{split}
\mathbf{F}&=
\begin{pmatrix}
\mathbf{F^{+}} & \mathbf{F^{\times}}
\end{pmatrix}\\
&=\begin{pmatrix}
F^{+}(t_0)e^{2\pi if_0 t_0+i\phi_{D}(t_0)} & F^{\times}(t_0)e^{2\pi if_0 t_0+i\phi_{D}(t_0)} \\
\vdots & \vdots \\
F^{+}(t_{n})e^{2\pi if_0 t_{n}+i\phi_{D}(t_{n})} &  F^{\times}(t_{n})e^{2\pi if_0 t_{n}+i\phi_{D}(t_{n})}
\end{pmatrix},
\end{split}
\end{equation}
$k=0,1,2, \ldots, N-1$ labels the data observed by the detector at the time $t_{k}=k *\Delta t$ and $1/\Delta t$ is the sampling rate.
The GW signal $\mathbf{s}=\mathbf{F}\mathbf{h}$ spanned by $\mathbf{F^{+}}$ and $\mathbf{F^{\times}}$ can be viewed as being in a
subspace of the space of detector output.
We can construct the \emph{null projector}  $\mathbf{P}_{\text{null}}(\hat{\omega}, f_0)$ \cite{Sutton:2009gi}
to project away the signal if the projector is constructed with the source's sky location and GW frequency \cite{Pang:2020pfz}. The null projector is given by
\begin{equation}
\mathbf{P}_{\text{null}}
=\mathbf{I}-\mathbf{F}(\mathbf{F}^{\dagger}\mathbf{F})^{-1}\mathbf{F}^{\dagger},
\label{eq:nullprojector}
\end{equation}
where $\dagger$ denotes the Hermitian conjugation.
Applying the null projector on the strain data $\mathbf{d}$ in Eq.~(\ref{eq:d_Fh}), we obtain
\begin{equation}
\begin{aligned}
\mathbf{z}&=\mathbf{P}_{\text{null}}(\hat{\omega},f_0)\mathbf{d}\\
&=\mathbf{P}_{\text{null}}(\hat{\omega},f_0)\mathbf{F}(\hat{\omega},f_0)\mathbf{h}
+\mathbf{P}_{\text{null}}(\hat{\omega},f_0)\mathbf{n}\\
&=\mathbf{P}_{\text{null}}(\hat{\omega},f_0)\mathbf{n},
\end{aligned}
\end{equation}
where $\mathbf{z}$ is the \textit{null stream} which only consists of the noise living
in a subspace that is orthogonal to the one spanned by $\mathbf{F}^{ +}$
and $\mathbf{F}^{ \times}$.

To consider the effect of polarization modes other than tensor modes, we parameterize the extra polarization modes in the tensor-scalar, tensor-vector and tensor-vector-scalar models with different strengths of non-tensorial components.
Following \cite{Chatziioannou:2012rf},
we take the waveforms of the extra polarization modes in the tensor-scalar model as
\begin{equation}\label{tensor_scalar}
\begin{split}
h_b=&B\mathcal{A}\sin^2{(\iota)}\exp{(2\pi i f_0 t + i\phi_0)},\\
h_l=&\sqrt{2}B\mathcal{A}\sin^2{(\iota)}\exp{(2\pi i f_0 t + i\phi_0)},
\end{split}
\end{equation}
where $B$ denotes the relative amplitude of the scalar modes to the tensor modes.
The waveforms of the extra polarization modes in the tensor-vector model are \cite{Chatziioannou:2012rf}
\begin{equation}\label{tensor_vector}
\begin{split}
h_x=&B\mathcal{A}\sin{(2\iota)}\exp{(2\pi i f_0 t + i\phi_0)},\\
h_y=&2B\mathcal{A}\sin{(\iota)}\exp{(2\pi i f_0 t + i\phi_0)},
\end{split}
\end{equation}
where $B$ denotes the relative amplitude of the vector modes to the tensor modes.
The waveforms of the extra polarization modes in the tensor-vector-scalar model are \cite{Chatziioannou:2012rf}
\begin{equation}\label{tensor_vector_scalar}
\begin{split}
h_x=&B\mathcal{A}\sin{(2\iota)}\exp{(2\pi i f_0 t + i\phi_0)},\\
h_y=&2B\mathcal{A}\sin{(\iota)}\exp{(2\pi i f_0 t + i\phi_0)},\\
h_b=&B\mathcal{A}\sin^2{(\iota)}\exp{(2\pi i f_0 t + i\phi_0)},\\
h_l=&\sqrt{2}B\mathcal{A}\sin^2{(\iota)}\exp{(2\pi i f_0 t + i\phi_0)},
\end{split}
\end{equation}
where $B$ denotes the relative amplitude of the non-tensorial modes to the tensor modes.
Including the polarization contents beyond tensor
polarizations, the signal can be written in the form
\begin{equation}
d(t)
=\sum_{A}\bar{h}_A F^{A}(\hat{\omega},f_0,t) e^{2\pi if_0 t+i\phi_D(t)}+ n(t).
\end{equation}
The observational data matrix \eqref{eq:d_Fh} becomes
\begin{equation}
\mathbf{d}[k]
= \textbf{F}^{t}(\hat{\omega},f_0) {\mathbf{h}}_{t}[k] + \textbf{F}^{e}(\hat{\omega},f_0) {\mathbf{h}}_e[k]
+ \mathbf{n}[k],
\end{equation}
where the superscript $t$ means summing over $+$ and $\times$, while the superscript $e$ means summing over whatever additional polarizations  present.
For example, the observational data matrix \eqref{eq:d_Fh} in the tensor-scalar model becomes
\begin{equation}
\mathbf{d}[k]
= \textbf{F}^{+}(\hat{\omega},f_0) {\bar{h}}_{+}[k]
+\textbf{F}^{\times}(\hat{\omega},f_0) {\bar{h}}_{\times}[k]
+\textbf{F}^{b}(\hat{\omega},f_0) {\bar{h}}_{b}[k]
+\textbf{F}^{l}(\hat{\omega},f_0) {\bar{h}}_{l}[k]
+ \mathbf{n}[k].
\end{equation}
The null stream obtained from the null projector with pure-tensor beam pattern matrix is given by
\begin{eqnarray}
\mathbf{z}[k]&=& \mathbf{P}_{\textrm{null}}(\hat{\omega},f_0)
\mathbf{d}[k] \nonumber\\
&=&\mathbf{P}_{\textrm{null}} (\hat{\omega},f_0)\mathbf{n}[k]
	+\mathbf{P}_{\textrm{null}}\textbf{F}^{e}(\hat{\omega},f_0)\mathbf{h}_{e}[k].
\label{Eextra}
\end{eqnarray}
The last term signifies the presence of extra polarizations other than tensor modes.
If there are additional polarization modes in GWs, then the data $\tilde{\mathbf{z}}[k]$ which is the discrete Fourier transformation of $\mathbf{z}[k]$ \cite{Sutton:2009gi} has a discrete component at $f_0$ in the frequency domain.

\subsection{Simulation Result}
\label{sec:3b}
\begin{figure}[htp]
\centering
  \includegraphics[width=0.9\columnwidth]{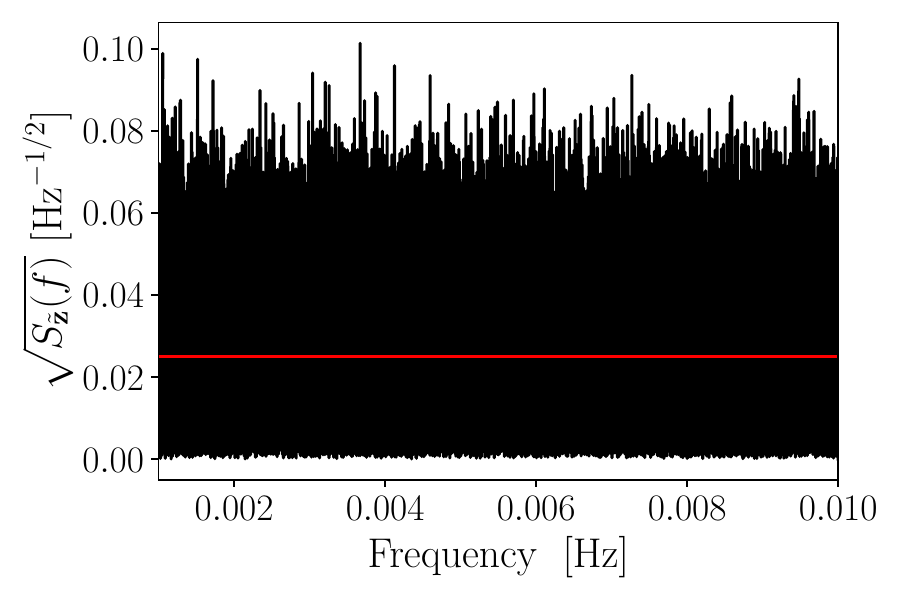}
\caption{The noise power spectrum $S_{ \tilde{\mathbf{z}} }(f)$. The red line is the average value of $S_{ \tilde{\mathbf{z}} }$.}
  \label{noisefig}
\end{figure}

For the reference source J0806.3+1527 and the total observation time of one year,
we choose the sampling rate as 0.02 Hz,
so $N=365\times 24\times 3600\times 0.02=630720$.
We inject a set of mock waveforms with $B=\{0,0.4,0.8\}$ in addition to simulated signals from GR.
The results for the tensor-scalar, tensor-vector and tensor-vector-scalar models with LISA,
TianQin and Taiji are shown in Figs. \ref{lisa_first_fig}, \ref{tq_first_fig}, and \ref{tj_first_fig} respectively.
The figures with $B=0$ show that this method can eliminate the tensor polarization if there is no extra polarization mode.
From figures with $B=\{0.4, 0.8\}$ we see that extra polarization modes in the tensor-vector and tensor-vector-scalar models can be  detected by LISA and Taiji.
For the tensor-scalar model, extra polarization modes with $B=0.8$ can be detected by Taiji.
The reason is that extra polarization signals should be loud enough for the detection.
The figures also show that extra polarization components with larger relative amplitude $B$ can be detected more easily.
From Fig. \ref{tq_first_fig}, we see that it is impossible to detect any extra polarization with TianQin for any model and any value of $B$ using this method.
To quantify the detection of extra polarization,
we use the signal-to-noise (SNR) $\rho$ \cite{Moore:2014lga},
\begin{equation}
    \rho^2=\int_{0}^{\infty}df \frac{4|\tilde{h}(f)|^2}{S_n(f)}= \frac{4|\tilde{\mathbf{z}}|_{f=f_0}^2 }{ S_{ \tilde{\mathbf{z}} } },
\end{equation}
where $|\tilde{\mathbf{z}}|_{f=f_0}$ represents the amplitude of $\tilde{\mathbf{z}}$ at frequency $f_0$ and $S_{ \tilde{\mathbf{z}} }$ represents the noise power spectrum of $\tilde{\mathbf{z}}$, shown in Fig.~\ref{noisefig}.
With $B=\{0.4, 0.8\}$ and one-year observation time, we get $\{\rho< 7,\rho< 7\}$ for LISA  and  $\{\rho<7,\rho=11.6\}$ for Taiji in the tensor-scalar model,
$\rho=\{57.4,113.3\}$ for LISA  and  $\rho=\{77.5,157.1\}$ for Taiji in the tensor-vector model,
$\rho=\{57.3,113.9\}$ for LISA  and  $\rho=\{77.6,154.9\}$ for Taiji in the tensor-vector-scalar model.
For TianQin we get $\rho<7$ in the tensor-scalar, tensor-vector and tensor-vector-scalar models with $B=\{0.4, 0.8\}$ and one-year observation time.
To get $\rho>7$ with one-year observation time, 
we find LISA requires $B>3.3$, $B>5\times 10^{-2}$ and $B>5\times 10^{-2}$
in the tensor-scalar, tensor-vector and tensor-vector-scalar models, respectively; 
Taiji requires $B>0.49$, $B>3.8\times10^{-2}$ and $B>3.8\times10^{-2}$ in the tensor-scalar, tensor-vector and tensor-vector-scalar models, respectively.
In order to avoid the limitation of the conclusion because it was drawn from one particular rather than a global representation of the performance of the detector,
we choose several representative locations for the J0806.3+1527-like source
and the results with LISA and Taiji are shown in Table \ref{snrtable}.
Except the location $(\theta, \phi)$, 
all other parameters for the sources are the same as J0806.3+1527.
For all the sources and models, we get $\rho<7$ with TianQin.
These results show that the conclusion that the method can be used by LISA and Taiji to detect extra polarizations is robust.
Due to the orbital motion of the detector in space,
along its trajectory, a detector like LISA and Taiji can be effectively regarded as a set of virtual detectors at different position and therefore form a network with $N$ number of virtual detectors to measure the polarization contents of monochromatic GW signals.
However, TianQin always points to the reference source J0806.3+1527 without changing the orientation of its detector plane,
so TianQin cannot use this method to detect the polarization contents of monochromatic GWs.

  \begin{widetext}
\begin{figure}[htp]
\centering
  \includegraphics[width=0.9\columnwidth]{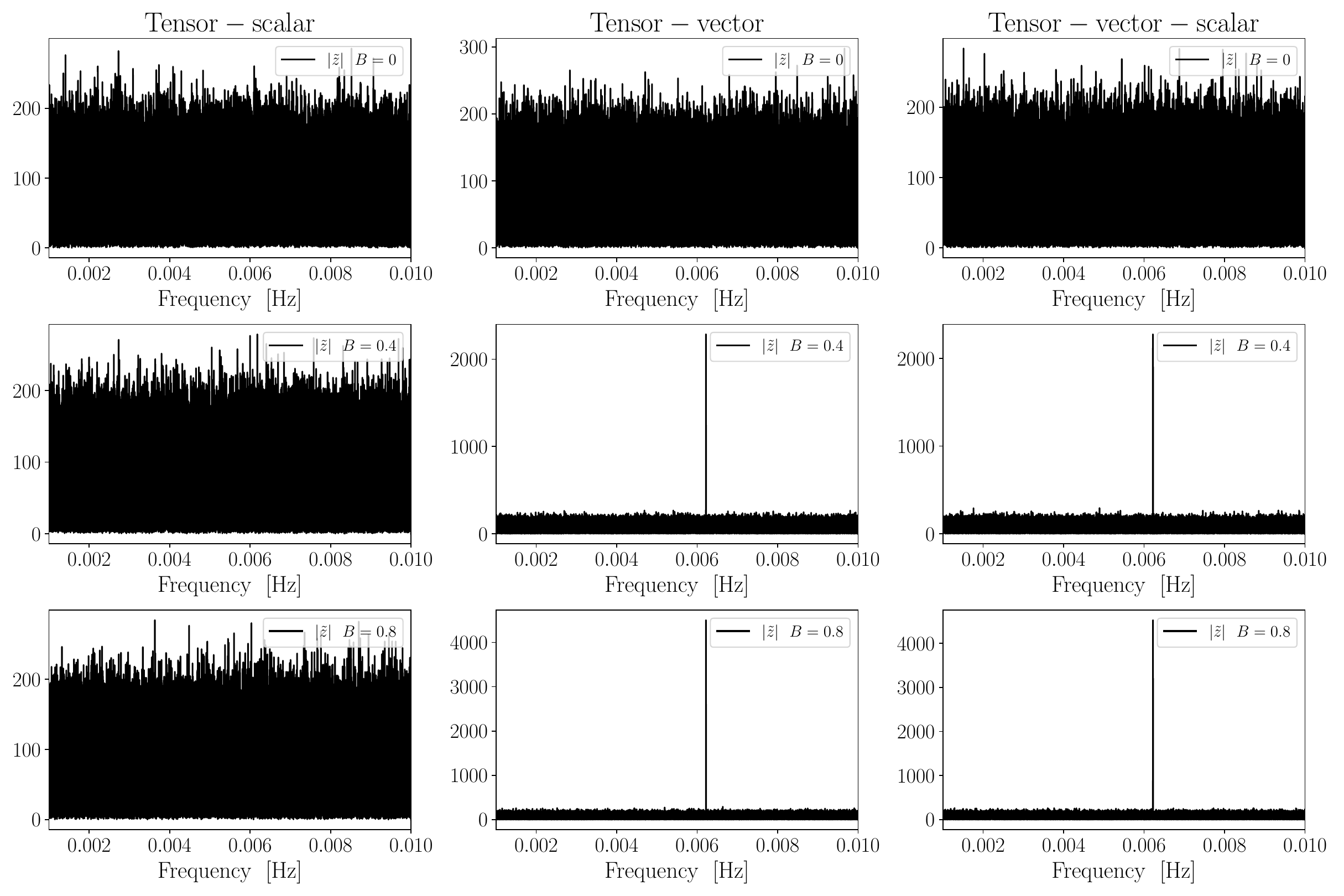}
\caption{The results of the null stream $|\tilde{z}|$  for the source J0806.3+1527 with LISA in the tensor-scalar (left panel), tensor-vector (middle panel) and tensor-vector-scalar (right panel) models.}
  \label{lisa_first_fig}
\end{figure}

\begin{figure}[htp]
\centering
  \includegraphics[width=0.9\columnwidth]{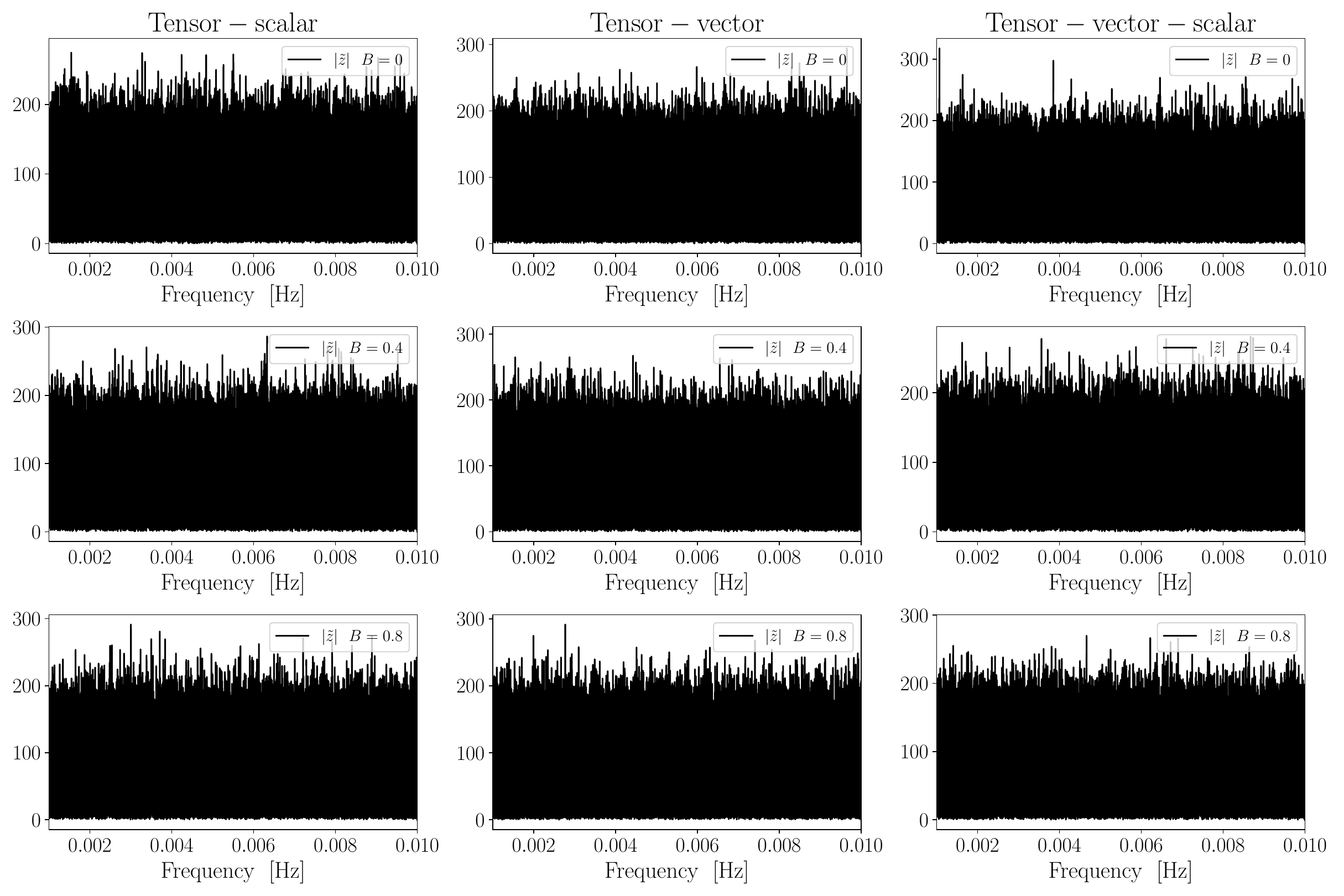}
\caption{The results of the null stream $|\tilde{z}|$  for the source J0806.3+1527 with TianQin in the tensor-scalar (left panel), tensor-vector (middle panel) and tensor-vector-scalar (right panel) models.}
  \label{tq_first_fig}
\end{figure}

\begin{figure}[htp]
\centering
  \includegraphics[width=0.9\columnwidth]{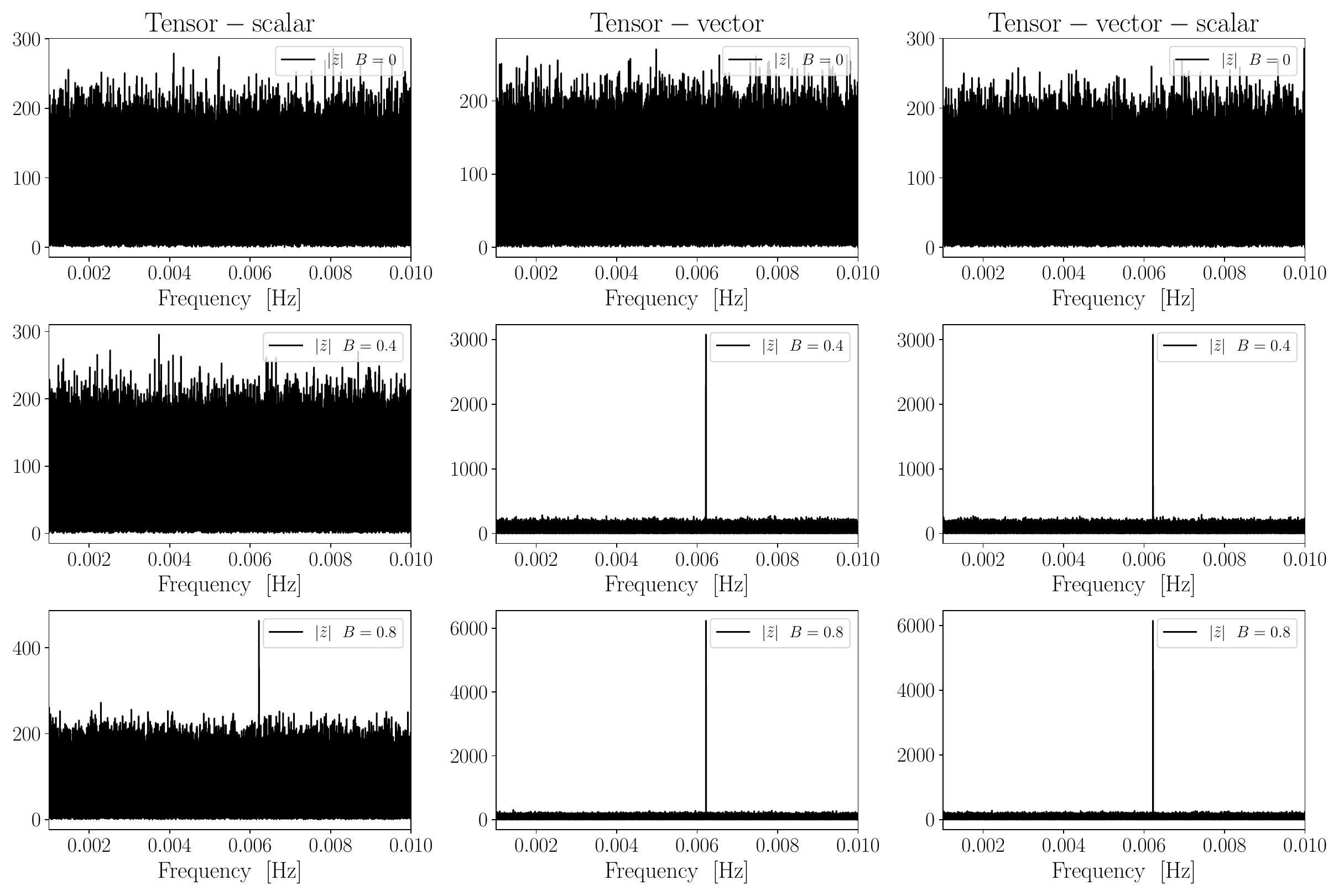}
\caption{The results of the null stream $|\tilde{z}|$  for the source J0806.3+1527 with Taiji in the tensor-scalar (left panel), tensor-vector (middle panel) and tensor-vector-scalar (right panel) models.}
  \label{tj_first_fig}
\end{figure}
 \end{widetext}

\begin{widetext}
 \begin{table}[htp]
  \centering
  	\begin{tabular}{|c|c|c|c|c|c|c|}
		\hline
 Location     &\multicolumn{2}{|c|}{Tensor-scalar model} &\multicolumn{2}{|c|}{Tensor-vector model}&\multicolumn{2}{|c|}{Tensor-vector-scalar model}\\ \hline
 ($\theta$, $\phi$) & LISA & Taiji & LISA & Taiji & LISA & Taiji  \\ \hline %
(0.3, 5.0)   & $\{<7,8.1\}$  & $\{8.4,13.5\}$ & $\{64.4,126.7\}$  & $\{93.0,188.3\}$ & $\{62.8,125.3\}$  & $\{98.2,191.4\}$  \\ \hline

(0.3, 1.0)    & $\{<7,<7\}$  & $\{8.3,15.6\}$ & $\{57.3,105.4\}$  & $\{133.3,263.0\}$ & $\{54.4,107.5\}$  & $\{126.0,257.4\}$ \\ \hline

(-0.3, 5.0)  & $\{<7,9.4\}$  & $\{7.4,11.7\}$ & $\{64.0,125.0\}$  & $\{90.7,181.7\}$ & $\{68.6,132.3\}$  & $\{87.4,174.1\}$ \\ \hline

(-0.3, 1.0)   & $\{<7,7.6\}$  & $\{<7,18.0\}$ & $\{55.7,111.7\}$  & $\{131.7,265.1\}$ & $\{51.6,109.5\}$  & $\{136.8,274.2\}$ \\ \hline

(1.0, 5.0)   & $\{<7,<7\}$  & $\{<7,9.9\}$ & $\{49.1,94.1\}$  & $\{79.7,159.5\}$ & $\{47.0,95.2\}$  & $\{81.4,166.4\}$ \\ \hline
	\end{tabular}
 \caption{
SNRs in LISA and Taiji for the tensor-scalar, tensor-vector and tensor-vector-scalar models with $B=\{0.4, 0.8\}$. In addition to the locations ($\theta, \phi$) listed in the table, the other parameters are ($\mathcal{M} = 0.3 ~M_{\odot}$, $D_L = 0.5$ kpc, $\iota = \pi/6$, $\phi _0$=0, $f_0=6.22 ~\rm{mHz}$).}
    \label{snrtable}
\end{table}
\end{widetext}

\begin{figure}[htp]
\centering
  \includegraphics[width=0.9\columnwidth]{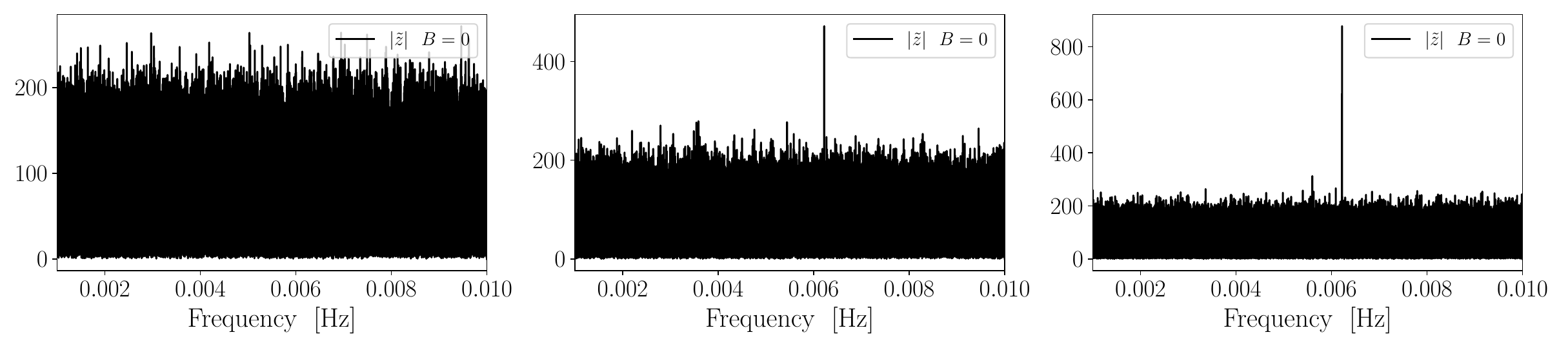}
\caption{The results of the null stream $|\tilde{z}|$  with the null projector constructed by a sky position different from the true source for LISA.
The sky position is at $(\theta=94.7^\circ+0.06^\circ, \phi=120.5^\circ+0.06^\circ)$ in the left panel, 
$(\theta=94.7^\circ+0.3^\circ, \phi=120.5^\circ+0.3^\circ)$ in the middle panel, 
and $(\theta=94.7^\circ+0.6^\circ, \phi=120.5^\circ+0.6^\circ)$ in the right panel.
}
\label{dsource}
\end{figure}

Accurately localizing GW sources is very important for measuring extra polarizations.
To show this point, we construct the null projector with a sky position different from the source's true location to project the signal and the results are shown in Fig \ref{dsource}.
From Fig.~\ref{dsource}, we see that when the sky position for constructing the null projector is away from the source's true location, 
the null projector can not eliminate the tensor polarizations.
In particular, if the localization error for the angles $\theta$ and $\phi$ is bigger than $0.06^\circ$, then the tensor signal cannot be eliminated from the data. 
Therefore, we can not distinguish extra polarizations from tensor polarizations if the sky location is not accurately known.
Fortunately, for space-based GW detectors, the accuracy of sky localizations is enough for constructing the null projector.

\subsection{IMPROVED METHODOLOGY}
The original methodology is based on the assumption that detectors like LISA and Taiji can be effectively regarded as a set of virtual detectors at different position and therefore form a network with $N$ number of virtual detectors to measure the polarization contents of monochromatic GW signals.
However, it is well known that instead of $N$ number of detectors, three detectors with different orientations are enough to discriminate extra polarization mode from the tensor modes.
Based on this fact, we split the data of one-year observation into three identical lengthy segments
with four-month data each and regard them as three independent detectors' data.
This improved method decreases the number of virtual detectors but increases the effective observation time for each detector.
It reduces computational memory and time because we only need to handle 3 dimensional matrix rather than $N$ dimensional matrix each time.
The observation time $T$ for each virtual detector becomes four months,
and the three data segments are
\begin{equation}\label{d0}
s_0(t)=\left(F^{+}(t)h_+(t) + F^{\times}(t)h_\times(t)\right)e^{i\phi_{D}(t)},
\end{equation}
\begin{equation}\label{d1}
\begin{split}
    s_1(t)=F^{+}&(t+T)h_+(t+T)e^{i\phi_{D}(t+T)} \\
    &+ F^{\times}(t+T)h_\times(t+T)e^{i\phi_{D}(t+T)},
\end{split}
\end{equation}
\begin{equation}\label{d2}
\begin{split}
    s_2(t)=F^{+}&(t+2T)h_+(t+2T)e^{i\phi_{D}(t+2T)} \\
    &+ F^{\times}(t+2T)h_\times(t+2T)e^{i\phi_{D}(t+2T)}.
\end{split}
\end{equation}
We rewrite the three detectors' observation data in the matrix form
\begin{equation}
\mathbf{d}(t) = \mathbf{F}(t)\mathbf{h}+\mathbf{n}(t),
\label{eq:d_Fh2}
\end{equation}
where
\begin{equation*}
\mathbf{d}(t)=
\begin{pmatrix}
d_{0}(t) \\
d_1(t) \\
d_{2}(t)
\end{pmatrix}
\text{,\,\,\,\,\,\,\,}
\mathbf{h}=
\begin{pmatrix}
\bar{h}_{+} \\
\bar{h}_{\times}
\end{pmatrix}
\text{,\,\,\,\,\,\,\,}
\mathbf{n}(t)=
\begin{pmatrix}
n(t) \\
n (t+T)\\
n(t+2T)
\end{pmatrix}
,
\end{equation*}
and
\begin{widetext}
\begin{equation}
\mathbf{F}(t)
=\begin{pmatrix}
F^{+}(t)e^{2\pi if t+i\phi_{D}(t)} &
F^{\times}(t)e^{2\pi if_0 t+i\phi_{D}(t)} \\
F^{+}(t+T)e^{2\pi if_0 (t+T)+i\phi_{D}(t+T)} &
F^{\times}(t+T)e^{2\pi if_0 (t+T)+i\phi_{D}(t+T)} \\
F^{+}(t+2T)e^{2\pi if_0 (t+2T)+i\phi_{D}(t+2T)} &
F^{\times}(t+2T)e^{2\pi if_0 (t+2T)+i\phi_{D}(t+2T)}
\end{pmatrix}.
\end{equation}
\end{widetext}
The signal can be seen as the data observed at a given time $t$ by three different detectors at the same time.
Within the observation period of four months, there are many observation points.
For any given time, we get
\begin{eqnarray}
\mathbf{z}(t) &=&\mathbf{P}_{\textrm{null}}(t)
\mathbf{d}(t) \nonumber\\
&=&\mathbf{P}_{\textrm{null}} \mathbf{n}
	+\mathbf{P}_{\textrm{null}}\textbf{F}^{e}(t)\mathbf{h}_{e},
\label{Eextra2}
\end{eqnarray}
where $\mathbf{z}(t)=(z_{0}(t),z_1(t),z_{2}(t))^T$.
For three virtual detectors, the total SNR is
\begin{equation}
    \rho^2=\rho_{\tilde{z}_0}^2+\rho_{\tilde{z}_1}^2+\rho_{\tilde{z}_2}^2.
\end{equation}
We apply the method \eqref{Eextra2} to detect extra polarizations in the tensor-scalar, tensor-vector and tensor-vector-scalar models.
For the reference source J0806.3+1527 and the total observation time of one year, we choose the sampling rate as 0.02 Hz.
The results are shown in Figs.  \ref{lisa_second_fig},  \ref{tq_second_fig}  and \ref{tj_second_fig} for LISA, TianQin and Taiji respectively.
For LISA and $B=\{0.4,0.8\}$, we get $\rho=\{148, 294\}$ in the tensor-vector model and $\rho=\{146,292\}$ in the tensor-vector-scalar model.
For Taiji and $B=\{0.4,0.8\}$, we get $\rho=\{19, 34\}$ in the tensor-scalar model, $\rho=\{258, 517\}$ in the tensor-vector model and $\rho=\{256,513\}$ in the tensor-vector-scalar model.
To get $\rho>7$ with one-year observation time, 
we find that LISA requires $B>3.1$, $B>1.9\times 10^{-2}$ and $B>1.9\times 10^{-2}$
in the tensor-scalar, tensor-vector and tensor-vector-scalar models, respectively;
Taiji requires $B>0.2$, $B>1.5\times10^{-2}$ and $B>1.5\times10^{-2}$ in the tensor-scalar, tensor-vector and tensor-vector-scalar models, respectively.

\begin{figure}
\centering
  \includegraphics[width=0.9\columnwidth]{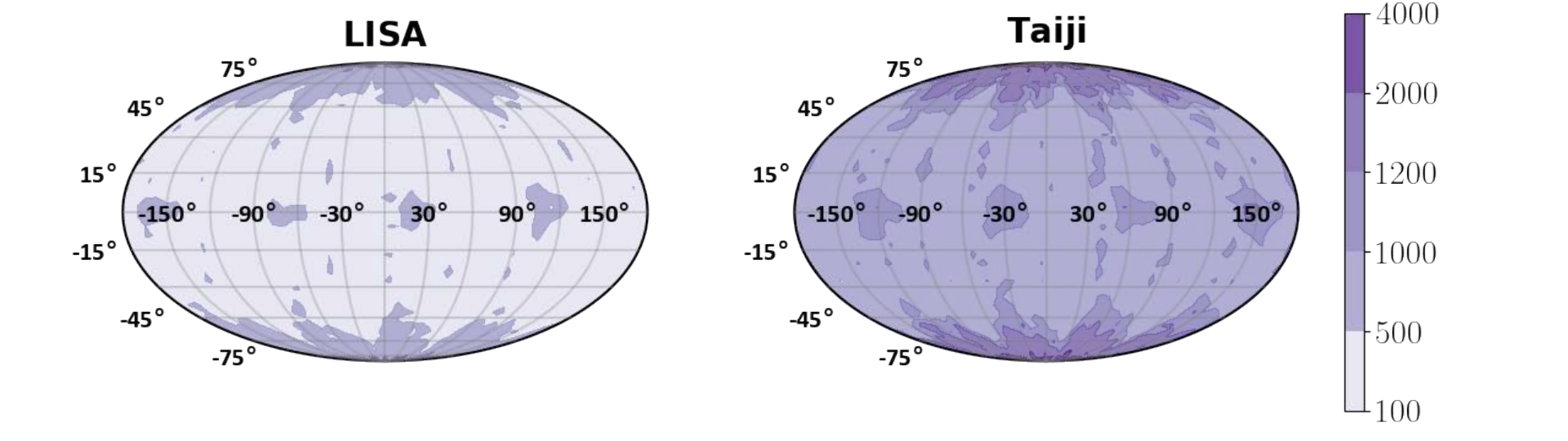}
\caption{
The sky map of SNRs in LISA (left panel) and Taiji (right panel) for the tensor-vector-scalar model with $B=0.8$.
	The parameters for the sources are $\mathcal{M} = 0.3 ~M_{\odot}$, $D_L = 0.5$ kpc, $\iota = \pi/6$, $\phi _0=0$ and $f_0=6.22 ~\rm{mHz}$.}
  \label{sky}
\end{figure}

\begin{figure}
  \centering
  \includegraphics[width=0.9\columnwidth]{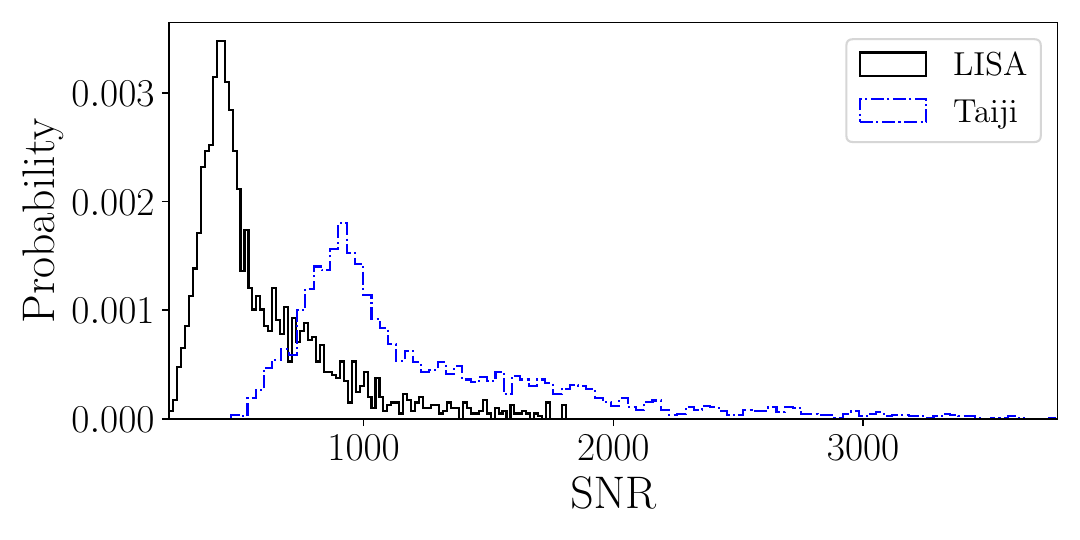}
  \caption{Histograms of SNRs  in LISA  and Taiji for the tensor-vector-scalar model with $B=0.8$.}
  \label{hist}
\end{figure}

We also simulate 2500 sources uniformly distributed in the sky with $-\pi/2<\theta<\pi/2$ and $-\pi<\phi<\pi$.
Except the locations, the other parameters of the sources are the same as the source J0806.3+1527.
Simulating the data in the detector with the waveforms
\eqref{tensor_vector_scalar} for the tensor-vector-scalar model with $B=0.8$,
we then apply the method \eqref{Eextra2} to calculate the total SNR. The sky map and the histogram of the SNR for Taiji and LISA are shown
in Fig. \ref{sky} and Fig. \ref{hist}, respectively.
The mean value of SNR is 571 with LISA and 1215 with Taiji for the tensor-vector-scalar model with $B=0.8$. The results show that LISA and Taiji can detect extra polarization modes with relative large $B$ for sources from all directions.

\begin{widetext}
\begin{figure}[htp]
\centering
  \includegraphics[width=0.9\columnwidth]{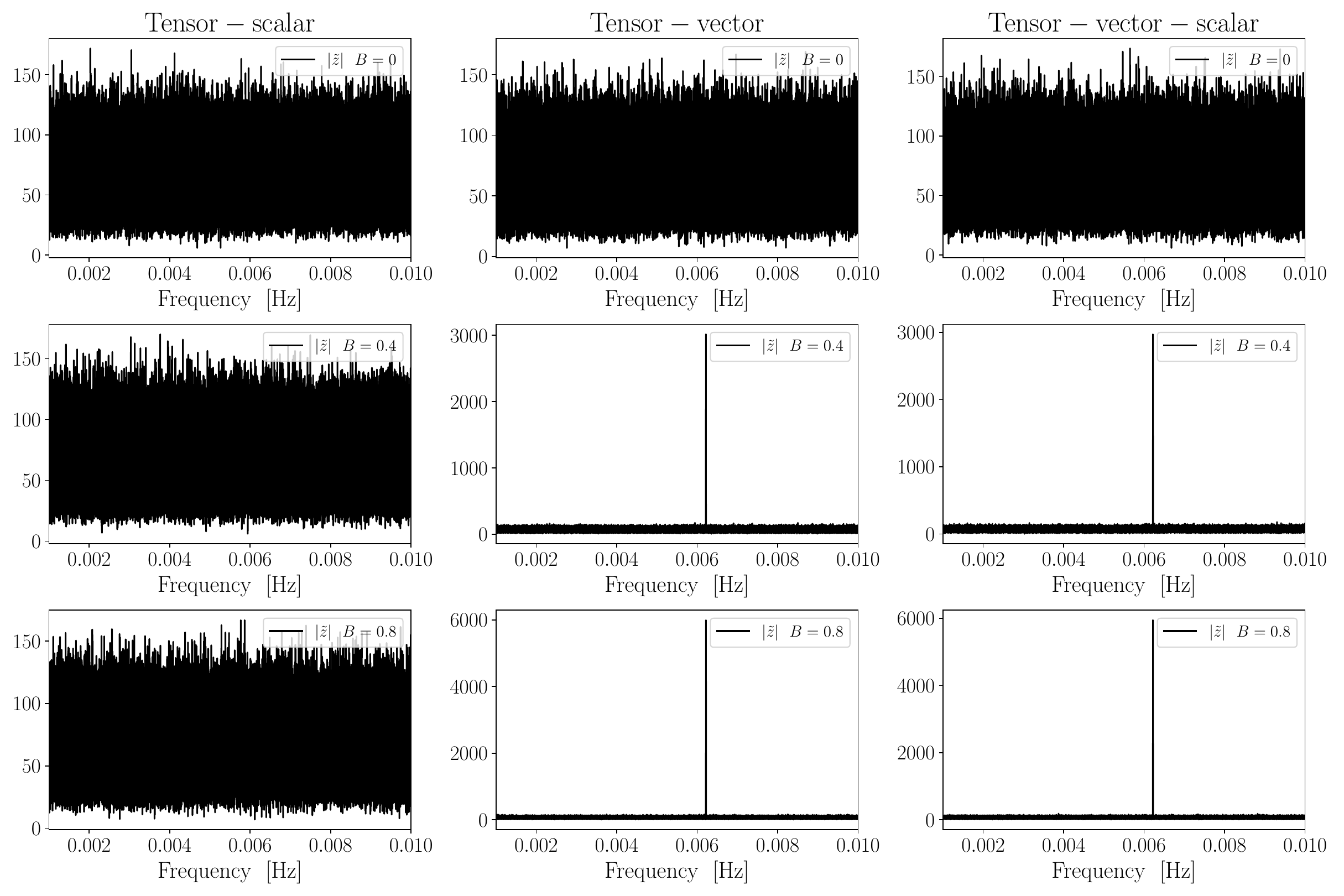}
\caption{The results of the null stream $|\tilde{z}|=\sqrt{|\tilde{z}_0|^2+|\tilde{z}_1|^2+|\tilde{z}_2|^2}$ for the source J0806.3+1527 with LISA in the tensor-scalar (left panel), tensor-vector (middle panel) and tensor-vector-scalar (right panel) models.}
  \label{lisa_second_fig}
\end{figure}

\begin{figure}[htp]
\centering
  \includegraphics[width=0.9\columnwidth]{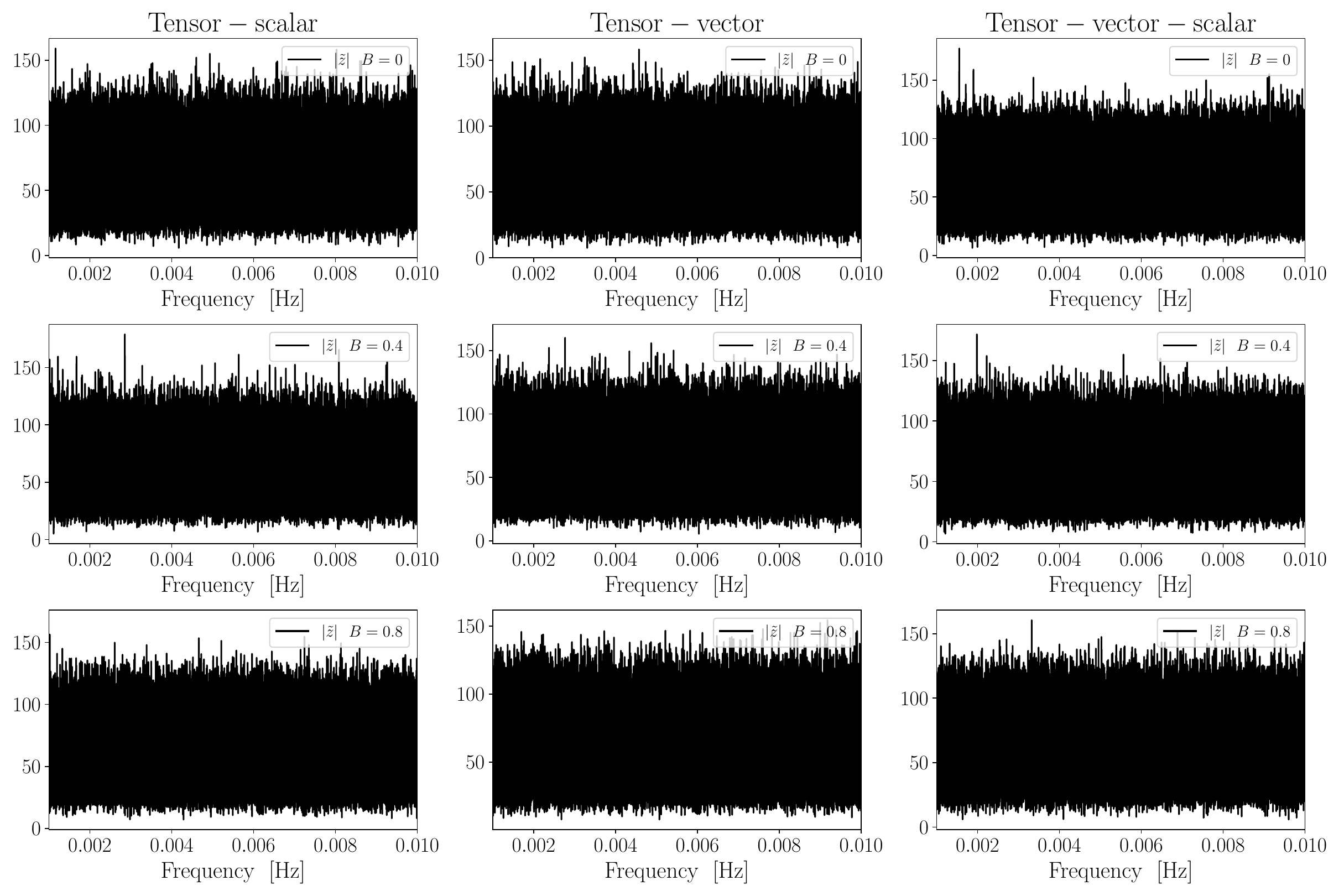}
\caption{The results of the null stream $|\tilde{z}|=\sqrt{|\tilde{z}_0|^2+|\tilde{z}_1|^2+|\tilde{z}_2|^2}$ for the source J0806.3+1527 with TianQin in the tensor-scalar (left panel), tensor-vector (middle panel) and tensor-vector-scalar (right panel) models.}
  \label{tq_second_fig}
\end{figure}

\begin{figure}[htp]
\centering
  \includegraphics[width=0.9\columnwidth]{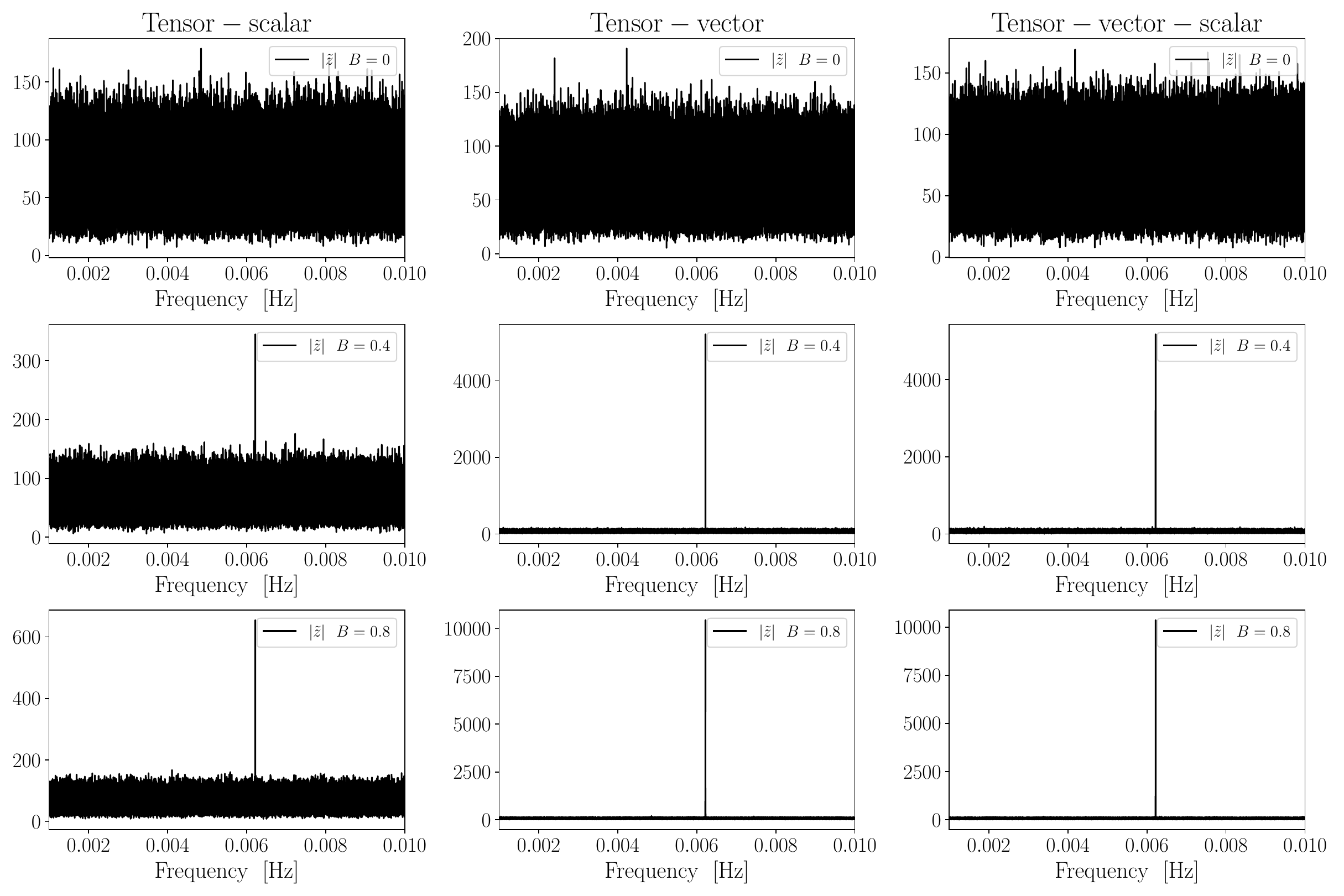}
\caption{The results of the null stream $|\tilde{z}|=\sqrt{|\tilde{z}_0|^2+|\tilde{z}_1|^2+|\tilde{z}_2|^2}$ for the source J0806.3+1527 with Taiji in the tensor-scalar (left panel), tensor-vector (middle panel) and tensor-vector-scalar (right panel) models.}
  \label{tj_second_fig}
\end{figure}
 \end{widetext}

\section{Conclusion}
\label{sec:final}
We introduce a concrete data analysis pipeline to test extra polarization modes of monochromatic GWs for space-based GW detectors.
This null stream method is applicable to LISA and Taiji because of their changing orientation of the detector plane.
We first take the single detector as $N$ virtual detectors by dividing the observational data into $N$ segments and use the source J0806.3+1527 as an example to simulate GW signals in the detector.
For one-year observation with the signal-to-noise of $\rho>7$,
we find that LISA can detect extra polarizations with the relative amplitude $B>3.3$, $B>5\times 10^{-2}$ and $B>5\times 10^{-2}$
in the tensor-scalar, tensor-vector and tensor-vector-scalar models, respectively;
and Taiji can detect extra polarizations with $B>0.49$, $B>3.8\times10^{-2}$ and $B>3.8\times10^{-2}$ in the tensor-scalar, tensor-vector and tensor-vector-scalar models, respectively.
We also analyzed the impact of the number of virtual detectors on the detection of extra polarization modes and we find that three virtual detectors with more observational time for each virtual detector can better detect extra polarization modes.

We then divide the one-year observational data into three identical segments to effectively form three virtual detectors.
With this method, the computational cost is much less.
For $\rho>7$,
LISA can detect extra polarizations with the relative amplitude $B>3.1$, $B>1.9\times 10^{-2}$ and $B>1.9\times 10^{-2}$
in the tensor-scalar, tensor-vector and tensor-vector-scalar models, respectively,
and Taiji can detect extra polarizations with $B>0.2$, $B>1.5\times10^{-2}$ and $B>1.5\times10^{-2}$ in the tensor-scalar, tensor-vector and tensor-vector-scalar models, respectively.
The results show that the ability of detecting extra polarizations is almost the same in the tensor-vector and tensor-vector-scalar models,
but the method is less effective in detecting extra scalar modes.
To discuss the dependence on the source location,
we simulate 2500 signals from the tensor-vector-scalar model with $B=0.8$ by distributing the sources uniformly in the sky,
the mean value of SNR is 571 for LISA and it is 1215 for Taiji.
If the sky location of the source is not accurately known,
then the method can not be applied to measure the polarizations. 
Therefore, this method can not be used to detect the polarization modes of stochastic GW backgrounds.
To detect the polarization modes of stochastic GW backgrounds, 
we need to combine multiple correlation signals as discussed in \cite{Nishizawa:2009jh}.
Similar to the idea of a virtual detector network considered in this paper,
the cross-correlation measured at different times can be regarded as
an independent set of signals with different location and separation,
these signals form a virtual network and help to improve the detection sensitivity \cite{Nishizawa:2009jh}.
By combining the technique of cross-correlation with our method, 
space-based GW detectors such LISA, TianQin and Taiji can detect polarization modes of stochastic GW backgrounds.

In conclusion, the method of the null stream can be applied to LISA and Taiji to detect extra polarization modes of monochromatic GWs.

\begin{acknowledgments}
This work is supported by
the National Natural Science Foundation of
China under Grant No. 11875136 and the Major Program of the National Natural Science
Foundation of China under Grant No. 11690021.
\end{acknowledgments}

\appendix
\section{DETECTOR'S ORBITS}
\label{orbits}
\subsection{TianQin's orbits}
In the heliocentric coordinate system,
the normal vector of TianQin's detector plane points to the direction of RX J0806.3+1527 with the latitude $\beta=94.7^\circ$ and the longitude $\alpha=120.5^\circ$. The orbits of the unit vectors of detector arms (two arms only) for TianQin are \cite{Hu:2018yqb}
\begin{equation*}
\label{detectorTQ}
\begin{aligned}
\hat{u}_x&=\cos(\omega_s t)\cos(\alpha)\cos(\beta)-\sin(\omega_st)\sin(\alpha),\\
\hat{u}_y&=   \cos(\alpha)\sin(\omega_st)+\cos(\omega_st)\cos(\beta)\sin(\alpha),\\
\hat{u}_z&=-\cos(\omega_st)\sin(\beta),\\
\hat{v}_x&=\cos(\omega_st+\frac{\pi}{3})\cos(\alpha)\cos(\beta)
-\sin(\omega_st+\frac{\pi}{3})\sin(\alpha),\\
\hat{v}_y&=   \cos(\alpha)\sin(\omega_st+\frac{\pi}{3})
+\cos(\omega_st+\frac{\pi}{3})\cos(\beta)\sin(\alpha),\\
\hat{v}_z&=-\cos(\omega_st+\frac{\pi}{3})\sin(\beta),\\
\end{aligned}
\end{equation*}
where the rotation frequency $\omega_s=2\pi/(3.65$ days).

\subsection{The orbits for LISA and Taiji}
In the heliocentric coordinate system, the detector's center-of-mass follows the trajectory
\begin{equation}
\bar{\theta}(t)=\pi/2,\quad \bar{\phi}(t)=2\pi t/T+\phi_\alpha,
\end{equation}
where $T$ equals one year and  $\phi_\alpha$ is just a constant that specifies the detector's location at the time $t=0$.
We set the initial phase $\phi_\alpha=-20^\circ$ for LISA and $\phi_\alpha=20^\circ$ for Taiji.
The orbits of the unit vectors of detector arms (two arms only) for LISA and Taiji are  \cite{Cutler:1998muh}
\begin{equation}\label{detectorLISA}
\begin{aligned}
\hat{u}_x&=-\sin(\bar{\phi}(t) ) \cos(\alpha_0(t)) \\
&\qquad\qquad+ \cos(\bar{\phi}(t)) \sin(\alpha_0(t))/2,             \\
\hat{u}_y&=     \cos(\bar{\phi}(t))\cos(\alpha_0(t))\\
&\qquad\qquad+\sin(\bar{\phi}(t))  \sin(\alpha_0(t))/2,        \\
\hat{u}_z&=\sin(\pi/3) \sin(\alpha_0(t)),         \\
\hat{v}_x&=-\sin(\bar{\phi}(t) ) \cos(\alpha_1(t)) \\
&\qquad\qquad+ \cos(\bar{\phi}(t)) \sin(\alpha_1(t))/2,             \\
\hat{v}_y&=     \cos(\bar{\phi}(t))\cos(\alpha_1(t))\\
&\qquad\qquad+\sin(\bar{\phi}(t))  \sin(\alpha_1(t))/2,        \\
\hat{v}_z&=\sin(\pi/3) \sin(\alpha_1(t)),         \\
\end{aligned}
\end{equation}
where $\alpha_i(t)$ increases linearly with time,
\begin{equation}
\alpha_i(t)=2\pi t/T-\pi/12-(i-1)\pi/3.
\end{equation}

\section{TDI for space-based GW antenna}
\label{tdigw}
\begin{figure}[htp]
	\centering
	\includegraphics[width=0.4\textwidth]{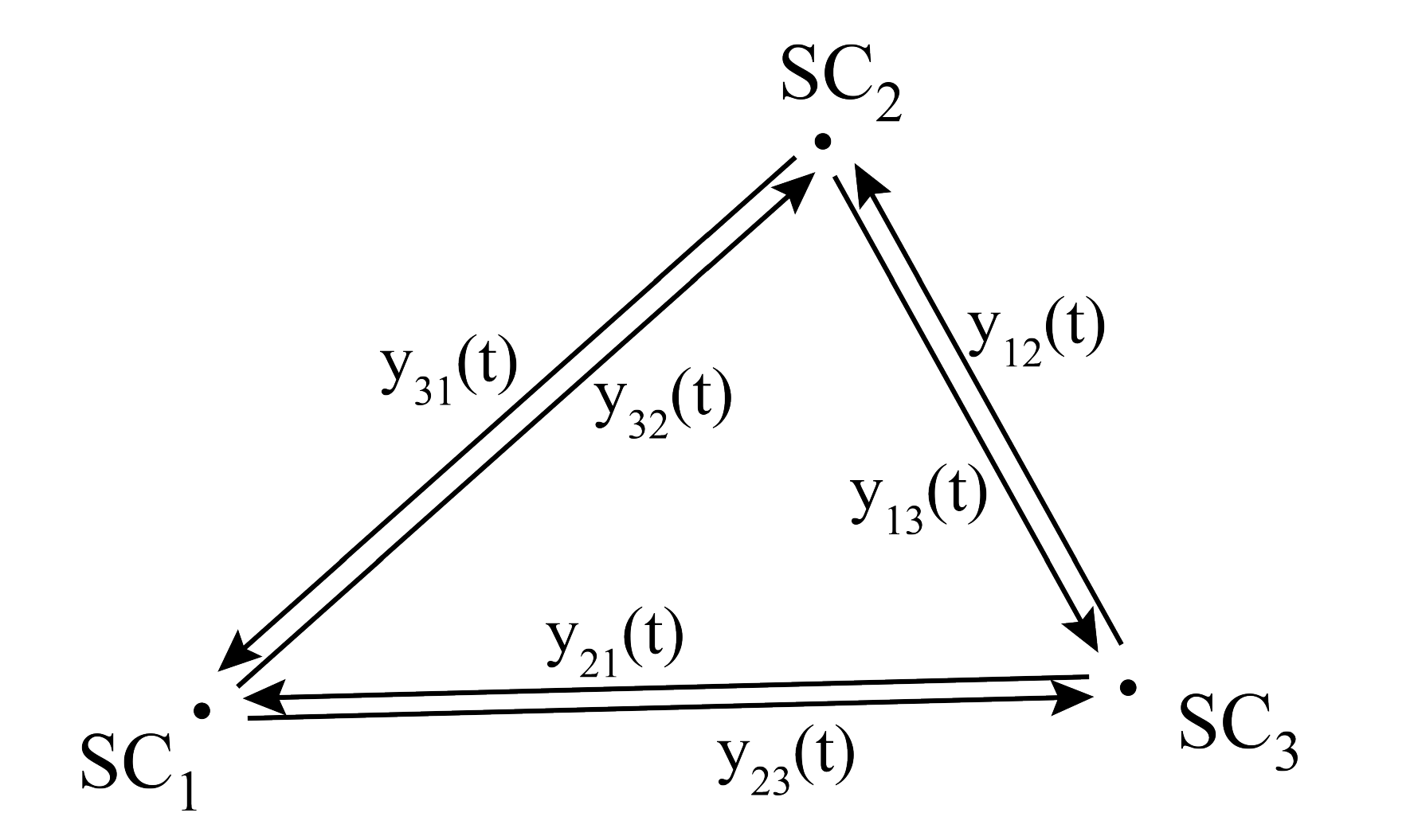}
	\caption{Six data beams $y_{ab}(t)$ exchanged between the spacecrafts.}
	\label{figdatabeams}
\end{figure}

\begin{figure}[htp]
\centering
  \includegraphics[width=0.9\columnwidth]{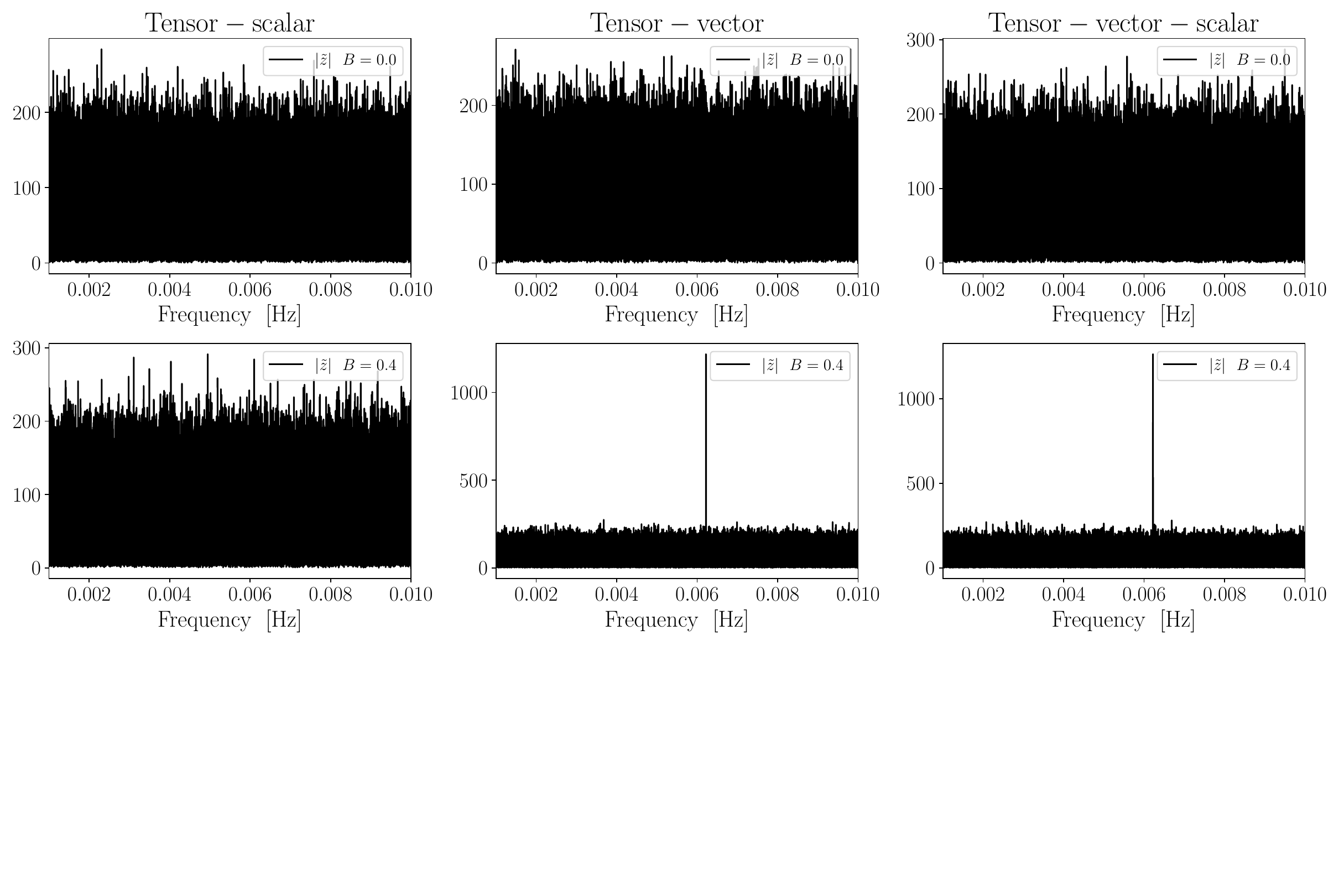}
\caption{The  TDI combination $X$ results of the null stream $|\tilde{z}|$  for the source J0806.3+1527 with LISA in the tensor-scalar (left panel), tensor-vector (middle panel) and tensor-vector-scalar (right panel) models.}
  \label{lisa_first_figtdi}
\end{figure}

Following \cite{Zhang:2019oet}, we show the relative frequency fluctuations time series $y_{ab}$ measured from detector $SC_d$ to detector $SC_b$ in Fig. \ref{figdatabeams}.
In the long-wavelength limit, we get the GW response for the six TDI signal
\begin{equation}
y_{ab}(t)=-\frac{1}{2}\sum_{ij} \hat{n}_{a}^i\hat{n}_{a}^j e_{ij}^Ah_A(t),
\end{equation}
where $\hat{n}_{a}$ is the unit vector along the arm.
The Michelson variable $X$ uses only four beams and two laser beams exchanged between two of the $SC_s$.
The GW response for $X$ is
\begin{equation}\label{tdix}
X=y_{32,322}-y_{23,233}+y_{31,22}-y_{21,33}+y_{23,2}-y_{32,3}+y_{21}-y_{31}+n(t),
\end{equation}
where the delayed data streams, e.g., $y_{23,2}=y_{23}(t-L_2)$, $y_{21,33}=y_{21}(t-L_3-L_3)$.
The noise in TDI combination $X$ is
\begin{equation}
\label{pnx}
S_n=[8\sin^2(4\pi f L)+32\sin^2(2\pi fL)]S_x/L^2+16\sin^2(2\pi fL)S_a(2\pi f)^{-4}/L^2,
\end{equation}
The signal in Eq.~\eqref{tdix} can be rewritten as
\begin{equation}
\begin{split}
d(t)=\bar{h}_+ &F^{+}(\hat{\omega},f_0,t) e^{2\pi if_0 t+i\phi_D(t)} \\
&+ \bar{h}_\times F^{\times}(\hat{\omega},f_0,t) e^{2\pi if_0 t+i\phi_D(t)}
+ n(t),
\end{split}
\end{equation}
where
\begin{align*}
\bar{h}_+=&\mathcal{A}\left[1+\cos^2(\iota)\right]\left[ e^{i\phi_0}+e^{i(\phi_0-2\pi f_0 L)}-e^{i(\phi_0-4\pi f_0 L)}-e^{i(\phi_0-6\pi f_0 L)} \right], \\
\bar{h}_\times=&2i\mathcal{A}\cos(\iota)\left[ e^{i\phi_0}+e^{i(\phi_0-2\pi f_0 L)}-e^{i(\phi_0-4\pi f_0 L)}-e^{i(\phi_0-6\pi f_0 L)} \right].
\end{align*}
Following the procedure discussed in Section \ref{sec:3b}, we inject a set of mock waveforms with $B=\{0,0.4\}$ in addition to simulated signals from GR.
The results for the tensor-scalar, tensor-vector and tensor-vector-scalar models with LISA are shown in Fig.
\ref{lisa_first_figtdi}.
With $B=0.4$ and one-year observation time, we get $\{\rho< 7,\rho=31,\rho=32\}$ for LISA in the tensor-scalar, tensor-vector and tensor-vector-scalar models, respectively.
These results are similar to those found in Section \ref{sec:3b}.

%

\end{document}